\documentclass[12pt,thmsa]{article}
\usepackage{amsmath}
\usepackage{graphicx}
\usepackage{amsfonts}
\usepackage{amssymb}

\begin{document}

\author{Arkady L.Kholodenko$\thanks{375 H.L.Hunter Laboratories,Clemson
University,Clemson,SC 29634-1905,USA. e-mail address:string@clemson.edu}$}
\title{Use of Meanders and Train Tracks for Description of Defects and Textures in
Liquid Crystals and 2+1 Gravity}
\date{}
\maketitle
\begin{abstract}
In this work the qualitative analysis of statics and dynamics of defects and
textures in liquid crystals is performed with help of meanders and train
tracks.It is argued that similar analysis can be applied to 2+1 gravity.More
rigorous mathematical justifications are presented in the companion paper
(Part II) on quadratic differentials and measured foliations. Meanders were
recently introduced by V.Arnold (Siberian J.of Mathematics \textbf{29},36
(1988)) and are used originally in the combinatorial problem of finding the
number of distinct ways given curve can intersect another curve in prescribed
number of points fixed along this auxiliary curve.Train tracks were introduced
by W.Thurston (Geometry and Topology of 3-Manifolds, Princeton U.Lecture
Notes,1979) in connection with description of homeomorphisms of two
dimensional surfaces.Train tracks alone are sufficient for the description of
statics and dynamics of liquid crystals and gravity .Using train tracks the
master equation is obtained which could be used alternatively to the
Wheeler-DeWitt equation for 2+1 gravity.Since solution of this equation is
possible but requires large scale numerical work, in this paper we resort to
the approximation of train tracks by the meanditic labyrinths. This then
allows us to analyse possible phases (and phase transitions) of gravity and
liquid crystals using Peierls-like arguments
\end{abstract}

\section{Introduction}

\subsection{ Motivations and background}

The role of topology in solving physical problems is steadily increasing [1,2
$].$In some instances it has become difficult to decide wether physical
arguments are helping to solve topological problems or topological arguments
are helping to solve physical problems $[2].$Here we would like to provide yet
another example of this sort.

In 3+1 dimensions the laws of Newton's gravity and Coulombic electrostatic
look suspiciously similar:both are being described by the Poisson-type equation%

\begin{equation}
\nabla^{2}\varphi=-\frac{4\pi}{\varepsilon}\rho\text{ ,} \tag{1.1}%
\end{equation}
where $\varepsilon$ is the dielectric constant in the case of electrostatics
while $\varepsilon^{-1}$=G is the gravitational constant in the case of
gravity.The density of charges $\rho(\mathbf{r)}$can be both positive and
negative for electrostatics while only positive in the case of gravity.The
solution of Eq.(1.1) for the potential $\varphi$(\textbf{r}) is given by%

\begin{equation}
\varphi(\mathbf{r})=\frac{1}{\varepsilon}\int\limits_{V}d^{3}r\frac
{\rho(\mathbf{r}^{\prime})}{\sqrt{(x-x^{\prime})^{2}+(y-y^{\prime}%
)^{2}+(z-z^{\prime})^{2}}} \tag{1.2}%
\end{equation}
where \textbf{r=}\{x,y,z\}and V is the volume which encloses the charges.It is
expected,that if the charges are confined within the domain V,then
$\varphi(\mathbf{r)\rightarrow}0$ when $\left|  \mathbf{r}\right|
\rightarrow\infty.$ Indeed, let $\rho(\mathbf{r})=q\delta(\mathbf{r}),$then we
obtain trivialy%

\begin{equation}
\varphi(\mathbf{r})=\frac{q}{\varepsilon}\frac{1}{\left|  \mathbf{r}\right|  }
\tag{1.3}%
\end{equation}
where q is the magnitude of charge placed at the orign.Solution given by
Eq.(1.2) is valid only in 3 dimensions,however.In two dimensions it should be
replaced by%

\begin{equation}
\varphi(\mathbf{r})=\frac{1}{\varepsilon}\int\limits_{A}d^{2}r^{^{\prime}}%
\rho\left(  \mathbf{r}^{\prime}\right)  \ln(\frac{1}{\sqrt{(x-x^{\prime}%
)^{2}+(y-y^{\prime})^{2}}}) \tag{1.4}%
\end{equation}
(with volume V being replaced by the area A) so that, instead of (1.3) ,we
obtain now%

\begin{equation}
\varphi(\mathbf{r)}=-\frac{q}{\varepsilon}\ln\left|  \mathbf{r}\right|  \text{
.} \tag{1.5}%
\end{equation}
This time,for $\left|  \mathbf{r}\right|  \rightarrow\infty$ we obtain
\quad$\varphi(\mathbf{r)\rightarrow\pm\infty}$ (depending upon the sign of
q).The result (1.5) can be obtained \textbf{directly }from Eq. (1.2) for a
special case of charges uniformly distributed along the infinitely long and
infinitely thin rod placed perpendicular to the x-y plane,i.e.%

\begin{equation}
-\frac{1}{2}\ln(\sqrt{(x-x^{\prime})^{2}+(y-y^{\prime})^{2}})=\lim
_{L\rightarrow\infty}(\frac{1}{2}\int\limits_{-L}^{L}dz^{\prime}\sqrt{\frac
{1}{(x-x^{\prime})^{2}+(y-y^{\prime})^{2}+(z-z^{\prime})^{2}}}-\ln2L)
\tag{1.6}%
\end{equation}
Instead of a rod we can imagine a particle ''living'' in 2+1
dimensions(i.e.particle which evolves in time while moving in x-y plane).

The difference of behaviors of potentials at infinity in two and three
dimensions leads to the profound difference in the underlying physics.
Moreover,this difference forbids smooth dimensional continuation of the
expression for the potential, given in d-dimensions by%

\begin{equation}
\varphi_{d}(\mathbf{r})=\frac{q}{\varepsilon}\frac{1}{\left|  \mathbf{r}%
\right|  ^{d-2}}, \tag{1.7}%
\end{equation}
based on the well known limiting procedure ,e.g. $\ln$x$=\lim\limits_{\alpha
\rightarrow0}\dfrac{x^{\alpha}-1}{\alpha}$ . Let us explain why this is
so.Since the potential $\varphi$ is an auxiliary quantity in both gravity and
electrostatics,we shall use instead the forces \textbf{F }obtained in a usual
way through%

\begin{equation}
\mathbf{F=}-\vec{\nabla}\varphi\text{ \quad.} \tag{1.8}%
\end{equation}
The forces create the vector fields and, in the case of 2 dimensions, we have
to consider the vector fields on surfaces.In particular,let us consider the
plane \textbf{R}$^{2},$the sphere S$^{2}=\mathbf{R}^{2}\cup\{\infty\}$ and the
disc D$^{2}$ of some radius R.Evidently,physically all three possibilities are
almost indistinguishable:the sphere S$^{2}$can be obtained from the plane by
one point compactification, the plane can be obtained from the disc D$^{2}$ by
taking the limit R$\rightarrow\infty$ .Topologically,however,the above tree
surfaces are completely different: the sphere S$^{2}$ has the Euler
characteristic $\chi(S^{2})=2$,the disc $\chi(D^{2})=1$ and the plane
\textbf{R}$^{2}$ cannot be given any value of $\chi$ [3]$.$ For the time being
we shall work with S$^{2}$ and D$^{2}$.The choice between these two surfaces
is determined by the boundary conditions.

Vector fields on the surface(manifold M) obey the Poincare-Hopf (P-H) index
theorem [4]:%

\begin{equation}
\chi(M)=\sum\limits_{i}I(x_{i}), \tag{1.9}%
\end{equation}
where the index I(x$_{i})$ of isolated singularity is defined as
follows.Consider the vector field \textbf{v}(x,y)=$\{$v$_{x}(x,y),$%
v$_{y}(x,y)\}$ around just one of the singularities and let $U\in M$ be some
(say ,circular) domain around this singularity. In this domain we can
construct a unit vector \textbf{n}(x,y) via%

\begin{equation}
\mathbf{n}(x,y)=\frac{\mathbf{v}(x,y)}{\left|  \mathbf{v}(x,y)\right|  }\text{
\quad.} \tag{1.10}%
\end{equation}
This vector provides the Gauss map from M to S$^{1}.$The $\deg$ree of this
mapping is I(x$_{i}).$More explicitly $[5,6],$%

\begin{equation}
I(x_{i})=\frac{1}{2\pi}\oint\limits_{C_{i}}\frac{\text{v}_{x}d\text{v}%
_{y}-\text{v}_{y}d\text{v}_{x}}{\text{v}_{x}^{2}+\text{v}_{y}^{2}}\text{
\quad.} \tag{1.11}%
\end{equation}
I cases of both Newtonian gravity and Coulombic electrostatics the index of an
isolated singularity is equal to one as can be easily seen by introducing the
complex variable z = x +y =re$^{i\varphi}$ and by combining Eqs (1.5),(1.8)
and (1.11).This means that the \textbf{minimal} number of charges which one
can put on the sphere is 2 while that on the disc is 1.The P-H
theorem,Eq.(1.9), \textbf{does not }automatically leads to the requirement of
electroneutrality: one can place two charges of the same sign on S$^{2}$
without violating this theorem.The complications arise,however, if one would
like to place \textbf{more than }2 charges on the sphere.In this case if
initially we would have,say,2 pluses,we would be unable to place the
additional charges while if we would have initially ''+'' and ''-''
charges,then such placement becomes possible \textbf{if we add charges
pairvise} while keeping the electroneutrality.This becomes possible only
because the presence of two identical charges creates a saddle (e.g.see Fig.1a
) which has an index -1,Ref.[4]. Hence,in the case of electrostatics the only
possible singularities are the sources/sinks (Fig.1d )with index 1 and the
\textbf{induced }saddles (Fig.1a)with index -1.\textbf{\ The existence of
multiple charges in electrostatics and the electroneutrality are directly
related to the topology of the underlying manifold and to the emergence of the
induced saddles.}

Consider now the case of Newton's gravity. Let us place the same 2 charges
(masses) on the sphere and let ,say,one of the charges be ''+'' so that
another is ''-'' and,since this is gravity ,the masses are naturally
attracting each other. Let us try now to add one additional mass (charge).We
immediately run into problem:since all masses attract each other,there cannot
be saddles. If this is the case,we are unable to place an additional mass on
S$^{2}$ .Hence,in such spherical universe we would have just 2 masses ! The
situation becomes even more dramatic if, instead of S$^{2}$ we would consider
D$^{2}$ .In this case we would not be able to put more than one mass while for
the case of an annulus(i.e.the disc with a hole)we would not be able to put on
it even a single mass! since the Euler characteristic of an annulus is 0,
Ref.[3]. At the same time,we can easily put e.g. 3 charges of the same sign on
the annulus in the case of electrostatics. These very simple arguments lead us
to the conclusion that there are some profound differences between the
Coulombic electrostatics and the Newtonian gravity and that the \textbf{above
differences alone are sufficient in order to arrive at the correct Einstein
formulation of gravity.}This is going to be demonstrated mainly in the
accompanying paper $[7]$ while here we provide only the qualitative arguments
$.$For this purpose we also need to discuss systems other than electrostatic
and gravitational (in Newton's sense).These are naturally occurring as defects
and textures in liquid crystals.

To study these defects it is helpful to recall some basic facts from the
qualitative theory of dynamical systems on surfaces[5,6].These are also
described in terms of flows of the vector fields.In two dimensions we are
usually dealing with the system of two equations%

\begin{align}
\frac{dx}{dt}  &  =\text{v}_{x}(x,y),\tag{1.12}\\
\frac{dy}{dt}  &  =\text{v}_{y}(x,y).\nonumber
\end{align}
Irrespective to the explicit form of the r.h.s. of Eq.(1.12) ,it is known
$[5,6]$ that the singularities of the flow could only be of the type depicted
on Fig.1\marginpar{Fig.1}\footnote{*Clearly,the flow directions on some of the
figures could be reversed}*.These can be easily obtained by linearization of
the Eq.(1.12) around the singularities of the vector field.The whole phase
portrait can then be built out of these local pictures by gluing together the
local pieces in some consistent way.Moreover,the topological considerations
allow us to restore the phase portrait with help of only \textbf{partial
}knowlege of the existing singularities.This principle is very helpful for our
case too but the situation is complicated by the fact that,in the case of
liquid crystals \textbf{as well as in }2+1 \textbf{gravity} ,instead of
\textbf{vector} fields on surfaces one has to deal with \textbf{line}
fields.This was demonstrated for liquid crystals in Refs.[ 10,11] and will be
proven for the case of gravity in Ref.[7] .The case of line fields has been
also studied in the theory of differential equations on surfaces$[5,6]$ .In
this case,instead of the system of Eqs (1.12),we have to consider%

\begin{equation}
\frac{dx}{\text{v}_{x}(x,y)}=\frac{dy}{\text{v}_{y}(x,y)}\text{ \quad.}
\tag{1.13}%
\end{equation}
Typical singularities of the line field are depicted in Fig.2 with the values
of the corresponding indices.\marginpar{Fig.2}For example,in the case of the
line fields with indices $\pm\frac{1}{2},$the differential equation which
describes these fields is given (in polar r,$\varphi$ coordinates)by$[5]$%

\begin{equation}
\frac{dr}{d\varphi}=r\tan(n\frac{\varphi}{2})\text{ \quad.} \tag{1.14}%
\end{equation}
For n=1 we obtain the line field with index -$\frac{1}{2}$ \quad which is
described analytically by%

\[
r=\frac{a}{\left(  \cos\frac{3\varphi}{2}\right)  ^{\frac{2}{3}}}%
\]
with $a$ being an arbitrary positive constant and $\varphi$ lying in the
following sectors

a) -$\frac{\pi}{3}<\varphi<\frac{\pi}{3}$ , b) $\frac{\pi}{3}<\varphi<\pi$ ,
c) $\pi<\varphi<\frac{5\pi}{3}$ .

Similar analysis can be performed for n=-1 which produces the index
\quad+$\frac{1}{2}$ as depicted in Fig.2. Although \textbf{for the line fields
it is impossible to introduce consistently the global orientation}$[4-6]$,it
is possible ,nevertheless, to introduce the indices of these fields $[4]$ by
analogy with the vector fields.The absence of orientation for the line fields
leads to the absence of ''forces'' between the singularities (defects). This
property is not unusual for the theories of 2+1 gravity where it is impossible
to introduce the interaction forces between the massive particles$[12]$. In
sections 4,5 of Part II (Ref.7) we shall demonstrate that both 2+1 gravity and
the textures in liquid crystals are described by the \textbf{line} fields. In
this part (which we call Part I )we shall rely mostly on the intuitive arguments.

Although it is impossible to introduce the interaction forces between defects
(or masses) it is possible,nevertheless,to talk about the total energy in both
2+1 gravity $[13]$ and in the theory of liquid crystals (e.g.see Part II and
section 5 below ).Moreover,because of this fact,it makes also sense to talk
about the different phases (orders) at least for the case of defects in liquid
crystals $[14]$ .Although the connections between the defects and textures in
liquid crystals and 2+1 gravity were discussed before $[15]$ , here we provide
completely different treatment of these connections.

\subsection{ Organization of the rest of this paper}

In section 2 we consider the role of topology in phase transitions in 2
dimensions.We argue,that in cases of the vector (Coulomb-like) and line fields
on surfaces topological considerations alone are sufficient for predicting the
Kosterlitz-Thouless type of phase transition$[14]$ from gas to dipole phase
while in case of the line fields in liquid crystals ,the existence of hexatic
phase $[16]$ can be easily established .Clearly ,topological considerations
alone provide only sufficient conditions for existence of ordered phases.The
necessary conditions require us to study the nature of the disordered phase in
some detail.This is accomplished in sections 3 and 4. In section 3 we provide
an explicit construction of the disordered phase using some ideas from the
qualitative theory of ordinary differential equations.Superposition of these
ideas with topological arguments produces notions of ''labyrinths'' and
''meanders'' .Since neither labyrinths nor meanders are widely known in
mathematical physics literature ,to avoid repetitions ,we refer the reader
directly to section 3 for precise definitions.Here ,we only would like to
mention that the description of labyrinths is closely related to the
description of the maps of the circle $[17,18]$ which should be more familiar
to physically trained reader.The notion of a meander has its origin in the
work by Arnold $[19]$ who actually invented this word. In the simplest case
one can think about the meander as a representative of the class of curves
which intersect another curve in the prescribed number of points.Although from
the above vague definition it may not be clear that the meanders and the
labyrinths may have many things in common,nevertheless,this \textbf{is }the
case as we demonstrate in section 3.

In section 4 we discuss yet another way of \quad looking at line fields
through the notion of \quad''train tracks'' . This way of looking at line
fields was proposed originally by Thurston $[20]$ in connection with his
studies of 3-manifolds. We argue in this section that the train tracks can be
used for description of dynamics of textures in liquid crystals and 2+1
gravity.The master equation which describes the evolution of the train track
could be used as an alternative to more familiar Wheeler-DeWitt equation (in
the case of gravity).Since the actual calculations which involve this equation
resemble that of the Heisenberg-type quantum mechanics,they require a large
scale numerical work which we try to avoid (at this stage of research) by
restricting ourself to the approximation of train tracks by meandritic
labyrinths. The legitimacy of this approximation is discussed in the Appendix.

Use of the meandritic labyrinth approximation allows us to obtain additional
quantitative information about the stability of the disordered phase and the
parameters of the order-disorder phase transition in the system of meandritic
labyrinths.This is accomplished in section 5.

All results of Part I are provided without serious proofs (with few
exceptions).Part II (Ref.[7]) serves to some extent to correct this deficiency
by providing the mathematical justifications to the emerging picture
.Sufficient details (and references) are provided in both parts of this work
to make this presentation self-contained and accessible not only to the
experts on gravity but to the interested condensed matter researchers as well.

\section{Role of topology in phase transitions in 2 dimensions}

We had already demonstrated (in section 1.1.) that the property of
electroneutrality is directly connected with the topology of the underlying
manifold, at least in the case of Coulombic -like systems. Here, we would like
to demonstrate that ,in addition,the topological considerations \textbf{alone
}determine the nature of phase transitions in such systems.For the line fields
the sequence of arguments ,unfortunately,is considerably more complicated and
will be discussed in the rest of this paper and in Part II.

In the case of Coulombic-like system, let us assume that under some favorable
conditions the incoming ''+'' and ''-'' have ''decided'' to stay together thus
forming a dipole.The vector field in this case is depicted in Fig.3 along with
a separate source (+) (or sink (-)).\marginpar{Fig 3.}The index of the dipole
is 2 so that P-H theorem is not violated if on the sphere we would have only 2
charges which have ''decided'' to form a dipole .Let us place a source/sink in
the vicinity of such dipole.By direct inspection we can notice that
topologically such source/sink cannot fit the vector field flow pattern coming
from the dipole unless it is combined with yet another charge thus forming yet
another dipole.Formation of the second dipole will cause the formation of 2
additional saddles so that P-H theorem is not violated. Evidently,this
procedure can be extended to other charges. From here several conclusions
could be drawn. First,the transition to the dipole phase should be sharp.
Second,since in the above arguments the density of charges $\rho$ was not
present, it means that such transition cannot be considered in traditional
thermodynamical sense ,that is for some critical temperature T$_{c}$ and
critical pressure P$_{c}$ there is no true critical density $\rho_{c}$ .These
conclusions are in complete accord with the results of Hague and Hemmer
$[21]$who had provided more traditional statistical mechanics treatment of
this type of transition.Obtained results are also in complete accord with the
results of more sophisticated treatment performed by Kosterlitz and Thouless $[22].$

The above simple picture breaks down immediately if we are willing to analyze
possible transitions in the case of defects which produce the line
fields,e.g.liquid crystals or 2+1 gravity.Unlike the Coulombic case where the
vector field picture is rather simple:or we have an independent charges or we
have dipoles, in the line fields case there are many more possibilities.Since
the local and global analysis of the line fields had been a subject of
intensive research in mathematics for quite some time ,it is impossible to
squeeze the enormous amount of results accumulated to date in this work. For a
good summary,please,consult Refs.[23-25 ].Without even trying to make a
summary of these results,we shall,nevertheless,select those which ,we feel
,are of immediate physical relevance.

Let us begin with the observation that in case of the Coulombic-like fields
the regrouping of charges had not produced a violation of the P-H theorem.
Evidently,we have to require the same in the case of line fields. The moves
which do not violate P-H theorem are depicted in Fig.4 and are known in the
literature as Whitehead moves[23]\marginpar{Fig.4} In addition,it is possible
to imagine the situations depicted in Fig.5 \marginpar{Fig.5}.Clearly,by
looking at Fig.4 we observe the case of \textbf{creation }of ''new '' defects
with index $\pm1$ from the ''old'' ones with index $\pm\frac{1}{2},$while
looking at Fig.5 we effectively observe the case of \textbf{destruction }of
defects. The indices of a) and b) are + $\frac{1}{2}$ while the index of c) is
zero. In principle,other situations are also possible but,for reasons which
will be explained in section 4 and Part II, we shall not be concerned here
with more complicated situations.

Given these \textbf{local} moves, what one can say about the ''phase
transitions'' (as compared to the Coulombic -like vector fields) in the system
of such defects ? For example,can we expect ,by analogy with Coulombic
case,that the transition(s) is ( are) going to be density-independent? For
this purpose let us consider the case of a disc $\chi(D^{2})=1$.To be in
accord with the P-H theorem, Eq.(1.9), we need to have at least four defects
with half integer indices (since only half integer types of defects are
topologically stable $[26])$ :one with index -$\frac{1}{2}$ and 3 with the
index +$\frac{1}{2}$ .The resulting ''stable phase'' is depicted in
Fig.6.\marginpar{Fig.6}This does not look like a dipole and,because of the P-H
theorem,we cannot just form another dipole and so on as in the Coulombic case
and we cannot use the electroneutrality condition as well. So,what else could
we imagine? For example,should we have \textbf{exactly }14 defects ,we could
have an orderly structure which resembles that depicted in Fig.7 .We use the
word ''periodic'' to describe this phase following Thurston's classification
of surface homeomorphisms $[25]$ as will be explained in more details in
section 4. \marginpar{Fig.7}

Is it possible to obtain something else ? The answer depends upon how many
defects are at our disposal.Consider the case of thermodynamic limit ,i.e.the
case when the size R of our disc D$^{2}$ is allowed to approach infinity .In
this situation we can imagine the structure depicted in
Fig.8.\marginpar{Fig.8}How can we check that this structure is actually in
agreement with the P-H theorem? This can be accomplished following the
original arguments by Hopf $^{4}($see also Ref..27). Let S be closed
orientable surface of genus g . Let a$_{0}$ be the number of vertices,a$_{1}$
be the number of edges and a$_{2}^{{}}$ be the number of 2-cells
(e.g.hexagons). Then,according to the Euler theorem,we obtain ,%

\begin{equation}
\text{a}_{0}-\text{a}_{1}+\text{a}_{2}=2-2\text{g} \tag{2.1}%
\end{equation}
Looking at Fig.8 and following the line of arguments made by Hopf,let us
replace the original hexagonal lattice with the elementary cell depicted in
Fig.9 \marginpar{Fig.9}

The face of each cell has an index +1,the edge has an index 2(-$\frac{1}%
{2})=-1$ and each vertex has an index 1.Whence,such construction holds for an
arbitrary g and,therefore,the hexagonal phase is consistent with the Euler
theorem ,Eq.(1.2) ,and,accordingly,with the P-H theorem,Eq.(1.9).As in the
case of the Kosterlitz and Thouless transition earlier described ,the
existence of the above hexatic phase is already known in physics
literature,e.g. see Ref. [28], where this phase is obtained based on
completely different set of arguments.In section 4 of Part II we shall provide
yet another arguments in support of the existence of this phase.

The questions arise:

a)is such obtaned hexatic phase unique ?

b)can such phase occur for an arbitrary concentration of defects (as in the

Coulombic phase case)?

The answer to the first question is negative as can be easily seen from the
situation depicted in Fig.10a) and the corresponding elementary triangle given
in Fig.10b).\marginpar{Fig.10}Taking into account that , in view of Fig.5c)
,the effective total index of singularities at the edge (a$_{0}$a$_{2}$) is
zero,we obtain the situation which coincides exactly with that described by
Hopf $[4]$ .That is we have the index of 2-cell the same as a$_{2}$ vertex
(i.e.+1),the index of the edge the same as the index at a$_{1}$ (i.e.-1) and
the index of the vertex is the same as the index at a$_{0}$ (i.e.+1).Hence,we
are again in accord with the Euler theorem, Eq.(2.1).

Since we have at least two types of hexagonal structures we may think about
some sort of phase transition between them.In addition,we have not actually
proved that only two hexagonal structures are possible .A similar but much
simpler problem of packing of hard hexagons on the triangular lattice was
considered by Baxter$[29]($ and, more recently ,by Monasson and Pouliquen$)$
with partial success (since the solution of the hard hexagon model is related
to the solution of 2d Ising model in the magnetic field which is not known in
general ).

Because of some similarities between the hard hexagons and the present liquid
crystal problem, it is useful to furnish some details since they will
eventually be used to provide an answer to the second question (about the
concentration dependence) posed above.To this purpose,in the case of hard
hexagons ,let us introduce the surface density $\rho$ via%

\begin{equation}
\rho=\frac{\text{n}}{\text{N}} \tag{2.2}%
\end{equation}
(a similar problem which involves the one component plasma was considered
recently in Refs [30,31]),where n is the total number of hexagons and N is the
total number of lattice sites (triangles).More useful quantity ,however,is the
packing fraction $\eta=3\rho$ .To understand the emerging factor of 3 , the
inspection of Fig.11 is helpful (see also Ref.32 )\marginpar{Fig.11}.By
construction,the complete packing corresponds to $\eta=1$ .Looking at Fig .10
,we easily can conclude that,in the case of the packing depicted in
fig.10a),the packing fraction is $\frac{1}{2}$ .Even in the case of hard
hexagons the packing problem is not completely under control.That is the
description of transition from one mode of partial packing to
another(e.g.loose $\eta<1$ versus dense $\eta=1)$ is still lacking $[29]$ .At
the same time,in the present case we are not even dealing with \textbf{hard
}hexagons.Our ''hexagons'' can be easily destroyed as it will be discussed in
section 6 of Part II.To analyse this more complicated situation,quantities
other than $\rho$ may be helpful. Following Ref.[30] ,let us introduce some
cut-off radius $r_{0}$ (the size of the disclination core $[14,26]$)as well as
the surface density $\hat{\rho}$ via%

\begin{equation}
\hat{\rho}=\frac{\hat{n}}{\text{A}}\text{ \quad.} \tag{2.3}%
\end{equation}
where A is the surface area and$\ \hat{n}$ is the total number of surface
defects (charges).Next,we can introduce the Wigner -Seitz radius $r_{w-s}$ via%

\begin{equation}
\pi r_{w-s}^{2}=\frac{1}{\hat{\rho}}\text{ .} \tag{2.4}%
\end{equation}
Finally,the filling fraction $\nu$ can now be introduced via%

\begin{equation}
\nu=\left(  \frac{r_{0}}{r_{w-s}}\right)  ^{2}=\pi r_{0}^{2}\hat{\rho}\text{
\quad.} \tag{2.5}%
\end{equation}
By construction, 0$\leq\nu\leq1$ so that $\nu$ can be used (instead of $\eta$
)to characterize the possible phase transitions.These results will be used in
section 5.

\section{Labyrinths and meanders}

The disordered phase of line field defects is no less interesting than the
ordered phase.Let us consider the simplest possible case of just four defects
in D$^{2}$ .The ''ordered'' phase is depicted in Fig.6 .To get a feeling of
the disordered phase,following Ref.[33], let us consider the ''foliation box''
$\mathcal{F}_{0}$ depicted in Fig.12a).\marginpar{Fig.12}The lines ,known in
mathematical literature as\textbf{\ foliations} (measured foliations to be
exact$[10,23-25])$ are free of singularities in the box a) and contain 2
singularities in the box b)(with account of the results depicted in
Fig.5c)).Following Ref.[33] ,we call these singularities ''Y'' and ''thorn''
respectively.Since the total index in the box b) is zero,we can
,evidently,place in the line field as many as we wish such Y-thorn doubles
without violating of the P-H theorem,Eq.(1.9).This gives us a certain freedom
of moving such objects around.In particular,if for the moment we would like to
forget about the ''tail '' of Y-singularity,then,we can concentrate our
attention at the foliations in semicircle(s).We shall restore the tail
afterwards ( to account for the P-H theorem).In the meantime,let us consider
collection of 3 foliated semicircles as depicted in Fig.13.\marginpar{Fig.13}

The foliations for S$_{1}$ are centered at a ,for S$_{2}$ at b,and for S$_{3}
$ at c respectively.The ratio of diameters D$_{S_{3}}/$D$_{S_{1}}=\alpha$ is
chosen to be some irrational number. Let $x$ denote the location of an
intersection of a given foliation leave with the axis [0,1]. Then,depending
upon the direction from which the foliation hits the x-axis ,one can
distinguish between the following 3 possibilities: 1)if the foliation hits
[0,1] by coming from S$_{1}($e.g.see point b),then let R$_{1}(x)=\alpha
-x,0\leq x\leq\alpha;$ 2)if the foliation hits [0,1] by coming from S$_{2}( $
e.g.see point c ) ,then let R$_{2}(x)=1-x;3)$ if the foliation hits [0,1]by
coming from S$_{3}$ (e.g.see point 2'), then let R$_{3}(x)=1-(x-\alpha
),\alpha\leq x\leq1.$ The meaning of thus introduced functions R$_{1},$%
R$_{2},$and R$_{3}$ becomes clear if we consider their compositions,e.g.
[R$_{1}\circ$ R$_{2}](x)$ and [R$_{3}\circ$ R$_{2}](x)$ . Looking at Fig.13
,we can notice ,that in both cases \quad we are dealing with the second return
map of the closed interval [0.1],e.g. see the paths 1-2-3 (for R$_{1}\circ$
R$_{2})$ and 1'-2'-3' (for R$_{3}\circ$R$_{2}$) . Explicitly,we obtain :%

\begin{equation}
\lbrack\text{R}_{1}\circ\text{R}_{2}](x)=\alpha-(1-x)\text{ ,} \tag{3.1}%
\end{equation}%

\begin{equation}
\lbrack\text{R}_{3}\circ\text{R}_{2}](x)=\alpha+1-(1-x)\text{ .} \tag{3.2}%
\end{equation}
If now we identify the ends of the interval ,i.e.0 and 1 ,then \textbf{both
}Eq.(3.1) and (3.2) are \textbf{translations }by $\alpha$ along the circle.The
theory of circle homeomorphisms is well documented $[17,18]$ and, for the sake
of continuity, we would like to remind to our readers several important facts
which will help us to clarify the results just obtained and those which will follow.

Let us notice that the circle S$^{1}$ is the quotient : S$^{1}=\mathbf{R}%
/\mathbf{Z}$ ,i.e.the circle is the factor group of real numbers
\textbf{R\ }modulo integers $\mathbf{Z}$ .Let $f:$ S$^{1}\rightarrow$S$^{1}$
be some circle homeomorphism and let $F:\mathbf{R}\rightarrow\mathbf{R}$ be a
lift of $f$ (e.g. $f($S$^{1})=\exp\{2\pi ix\}$ and $F$ : $x^{\prime}=x+\alpha$
,that is $F$ are translations in \textbf{R}).The rotation number $\tau(F)$ can
be defined via equation :%

\begin{equation}
\tau(F)=\lim_{\left|  n\right|  \rightarrow\infty}\frac{1}{n}(F^{n}%
(x)-x)\text{ .} \tag{3.3}%
\end{equation}
To define $F(x)$ we notice that ,in general,%

\begin{equation}
F(x)=x+a(x),\qquad a(x+1)=a(x),\qquad0\leq x<1\text{ \quad\quad.} \tag{3.4}%
\end{equation}
In terms of $a(x)$ the rotation number can be written as well as%

\begin{equation}
\tau(F)=\lim_{\left|  n\right|  \rightarrow\infty}\frac{1}{n}%
(a(x)+a(F(x))+...+a(F^{n-1}(x)))\equiv\lim_{\left|  n\right|  \rightarrow
\infty}\frac{1}{n}a_{n}(x)\text{ .} \tag{3.5}%
\end{equation}
Let the mapping $F$ has a fixed point $x^{*}$ \quad for some q:%

\begin{equation}
F^{q}(x^{*})=x^{*}+a_{q}(x^{*})\text{ .} \tag{3.6}%
\end{equation}
It can be shown that,%

\begin{equation}
a_{q}(x^{\ast})=p \tag{3.7}%
\end{equation}
and
\begin{equation}
\tau(F)=\frac{p}{q}\text{ .} \tag{3.8}%
\end{equation}
That is q translations in the lifting space are equivalent to ''pure''
rotation of the circle by the angle $\vartheta=2\pi\frac{p}{q}$ .Accordingly
,if Eq.(3.6) \textbf{does not} have fixed point(s) for \textbf{any }q,then
$\tau(F)=\alpha$ where $\alpha$ is irrational.With respect to the situatioin
depicted in Fig.13 we have to conclude ,that the ''motion'' along the line
1-2-3,etc will never stop or cross another line,i.e. we have obtained the
foliation which is actually a labyrinth. Let us take now into account the P-H
theorem and make the situation more realistic.To this purpose we attach the
''tail'' to S$^{1}$semicircle and gently open the foliation leaf starting at
$\alpha$ in Fig.13. The result of such opening procedure is depicted in Fig.14
\marginpar{Fig.14}which represents the simplest example of a labyrinth: all
incoming leaves are being trapped inside the labyrinth forever.

The picture just described can be wasty generalized with help of the following
auxiliary observation.Consider some nonintersecting line $\mathcal{L}$ (closed
or not) and place on the top of it a finite collection of foliated half
circles(semicircles) provided that none of these half circles are touching
each other (unless the otherwise is specified).Let us place the remaining half
discs on the bottom of $\mathcal{L}$ in the way depicted in
Fig.15.\marginpar{Fig.15} We have obtained in this way an example of a
''meander'' labyrinth.Let us explain now what actually the word ''meander''
means.According to Ref.[34] ,a meander of order n is a closed
nonselfintersecting curve which intersects another straight line in exactly 2n
preassigned points(more exactly,an equivalence class of such closed curves
which leave the straight line $\mathcal{L}$ fixed).Evidently ,the line
$\mathcal{L}$ need not be straight in general.A few examples of the meanders
of lower order and the corresponding meandritic labyrinths are depicted in
Fig.16.\marginpar{Fig.16}Although the four last meandritic labyrinths are
degenerate (two half discs are touching each other),this may be still
permissible (if other discs are not touching each other),e.g.see Example 2 of
Ref.[33] .This is so because we still have to attach ''tails'' to some of the
discs (e.g. see Fig.14) to make them Y-type in order to be in accord with P-H
theorem.Trough the tails the external lines penetrate into the labyrinth and
become trapped forever.Moreover,because of the tails,the symmetry between
different labyrinths depicted in Fig.16 becomes broken so they \textbf{all
}have to be considered,just like the underlying meanders.

The meandritic number M$_{n}$ (that is the total number of meanders of order
n) provides the degeneracy factor for the partition function of defects to be
discussed in section 5. Its value is essential in the description of the
diclination binding-unbinding ''melting'' transition $[14]$in the non
-Coulombic two dimensional ''plasma'' of such defects. Actually,the above
description is still incomplete since ,as it is well known from statistical
mechanics of two dimensional plasma$[21]$ ,the clustering effects should also
be taken into account.This ,in particular,means that,instead of just one
connected meander,we may as well consider a collection of non-intersecting
meanders. Specifically,for the multicomponent meander of order n we may choose
k nonintersecting lines (1$\leq$k$\leq$n) so that the total number of
crossings on \textbf{all} lines is 2n. Accordingly,one can introduce the
meandritic number M$_{n}^{(k)}$ to account for this fact.$[35,36]$ .This
number will be estimated and used in section 5.In the meantime,it is useful to
introduce ,yet another,set of meanders,the projective meanders $[19,37]$
.These are defined as follows.Consider 2n points on the circle S$^{1}$.Divide
the points into n pairs so that the chords connecting points in each pair will
not intersect.Identify the diametrically opposite points on the circle,thus
turning the disc into the projective plane(or sphere).Such formed set of
curves in the disc is called the projective meander of order n.Evidently,the
number of possible projective meanders of order n is equal to the n-th Catalan number%

\begin{equation}
C_{n}=\frac{(2n)!}{(n+1)!n!} \tag{3.9}%
\end{equation}
which obviously reflects the combinatorics of the problem.We shall need the
notion of the projective meanders in order to demonstrate that the above
labyrinth construction is not artificial but ,actually ,is intrinsic for the
whole description of the measured foliations on surfaces.In addition,there is
yet another way to look at the whole problem for the case of line fields.
Because of its potential usefulness for the problems which involve 2+1 gravity
,we discuss this other approach in the next section.

\section{Train tracks and pseudo-Anosov homeomorphisms}

The logical development of dynamical systems on surfaces goes from the
detailed treatment of the circle maps through the torus maps and then, to
consideration of flows (foliations)on the Riemann surfaces of genus higher
than one .Nevertheless,already the circle maps exhibit all qualitative
features of more complicated situations.Moreover,as it was shown by Thurston
$[38]$ ,\textbf{all }surface homeomorphisms can be actually related to the
circle maps through the construction which he calls the ''earthquake''.

Let us briefly explain why this is so.(In the Appendix of Part II these
arguments will be extended to 3 dimensional hyperbolic manifolds).As in the
case of the circle maps discussed in the previous section,the surface
homeomorphism $\quad f:$ $R\rightarrow R^{\prime}$ (where both $R$ and
$R^{\prime}$ are some Riemann surfaces) can be lifted to the universal
covering space which is either the Poincare upper halfplane H defined by%

\begin{equation}
\text{H=\{z=x+iy}\in C\mid y>0\} \tag{4.1}%
\end{equation}
or the open unit disc D$^{2}$%

\begin{equation}
\text{D}^{2}=\{\text{w=u+iv }\in C\mid\text{u}^{2}+\text{v}^{2}<1\} \tag{4.2}%
\end{equation}
which is related to H via mapping%

\begin{equation}
\text{w=}\frac{\text{z-i}}{\text{z+i}}\text{ \quad, z}\in\text{H \quad.}
\tag{4.3}%
\end{equation}
Since every Riemann surface $R$ of genus g greater than one can be obtained as
the quotient $R=$H/$\Gamma$ where $\Gamma$ is some Fuchsian group,i.e.the
group of linear fractional transformations of the type%

\begin{equation}
\gamma(z)=\frac{az+b}{cz+d}\text{ , }ad-bc=1 \tag{4.4}%
\end{equation}
with $a,b,c,d$ being \textbf{real }numbers ,we can always lift the flow on
surface $R$ to the flow on D$^{2}$ where by means of earhquakes the
description for flows becomes connected with the description associated with
maps of the circle.Let now $\mathcal{C}$ be some closed curve on $R$ which is
homotopically nontrivial(that is it cannot be shrunk to the point).It can be
proven $^{39}$ that,

\textbf{Theorem 4.1.}$\quad C$\textit{\ is freely homotopic to a unique closed
geodesic l .}

Also,it is known $[40]$ that,

\textbf{Theorem 4.2}. \textit{For a closed Riemann surface of genus g there
are 3g-3 independent closed geodesics (please,see Fig.5 of Part II ).}

Finally,following Thurston $[25]$ and Casson and Bleiler $[41]$ ,we introduce

\textbf{Definition 4.1.} \textit{Disjoint union of geodesics is called
lamination(}$L)$\textit{\ ;the geodesics contained in }$L$\textit{\ are called
leaves of }$L$\textit{\ .}

It can be very easily shown $[17,18]$ that,when lifted to H, the geodesics
look like that depicted in Fig.17.That is they are either halfcircles with
center on the real axis or the lines perpendicular to the real axis
\marginpar{Fig.17}.When transformed from H to D$^{2}$ ,the halfcircles go to
halfcircles and the vertical lines to the diameters of D$^{2}$ as depicted in
Fig.18 .\marginpar{Fig.18}If we consider the case of lamination $\mathcal{L}$
then,by construction,the geodesics are nonintersecting on $R$ and,whence on
D$^{2}[$42]$.$ This means that \textbf{in }D$^{2}$ \textbf{they look like the
projective meanders}(before identification of points on S$^{1})$ described in
the previous section.A simple minded permutation of these geodesics in D$^{2}$
produces the Catalan number $C_{n}$ ,Eq.(3.9). In general,the situation is
considerably more complicated $[25,41]$ .

If for a given Riemann surface of genus g with n punctures we have the
inequality 2g+n-2
$>$%
0,then the mapping class group $M$($R)$ is defined through its action on
$\mathcal{L}$ .If $f\in M(R)$ and $C=\{c_{1},...,c_{r}\}\in\mathcal{L}$ ,then
the mapping $f$ is \textbf{reducible} if%

\begin{equation}
f(C)=C \tag{4.5}%
\end{equation}
Alternatively,it is being said that $f$ is \textbf{reduced }by $C.$

If, in accord with the theory of braids $[43]$ ,we assume that Eq.(4.5) holds
for any permutation of the set $C$ (and this is indeed the case $[41])$ ,then
the Catalan number ,Eq.(3.9) ,naturally emerges .The reducible case is not the
only possibility however.If,say, $f$ is reduced by $C$ ,then let $S=\{s_{1}$
,...$,s_{m}\}$ be the set of components of the complement of $R-C$ e.g.see
Fig. 5 of Part II . Since $f$ permutes $s_{i}$ ,let $n_{i}$ be the smallest
power of $f$ such that%

\begin{equation}
f^{n_{i}}(s_{i})=s_{i}\text{ .} \tag{4.6}%
\end{equation}
For any $n_{i}>0$ such mapping is called \textbf{periodic }.It could as well
be that Eq.(4.6) does not have fixed point(s) for any finite $n_{i}.$ In this
case,if Eq.(4.5) still holds,then the mapping is still reducible , but if
neither (4.5) or (4.6) hold , then the mapping is called
\textbf{pseudo-Anosov.} This term will be explained below.Before doing so,we
notice that Eq.(4.5) can be lifted to D$^{2}$ so that with respect to the ends
of geodesics lying on S$^{1}$ we have some sort of a mapping of a circle
analogous to that discussed in the previous section.According to Thurston
$[38]$ ,''every homeomorphism of the circle S$^{1}$ extends to the
homeomorphism of the disc D$^{2}$ ''.(In the Appendix to Part II we shall
discuss the extension of this result to the case of 3-manifolds). Evidently,
the reverse should be true as well.In which case , the lift of Eq.(4.6) to
D$^{2}$ has its analogue in Eq.(3.6) and the case when Eq.(3.6) does not have
solution for \textbf{any }\quad q corresponds to the pseudo -Anosov type of
mapping. Let us now explain the meaning of the word ''pseudo-Anosov''.
According to Refs.[17,18] ,the following theorem can be proven.

\textbf{Theorem} \textbf{4.3}.\textit{The geodesic flow on H/}$\Gamma
$\textit{\ is the Anosov flow}.

The best way to explain this is again pictorial .We lift the geodesics from
$R$ to H in order to obtain the picture given in Fig.19.\marginpar{Fig.19}%
Accordingly,we can also look at the same picture by using the disc D$^{2}$
model.In this case we have the situation depicted in Fig.20.\marginpar{Fig.20}%
The circles tangent to x-axis at some point x$_{0}$ (in H-plane) are called
\textbf{horocycles.}It can be shown,that the set of horocycles (touching
x-axis at x$_{0})$ and the set of geodesics (emanating from x$_{0}^{{}})$ are
mutually orthogonal (thus forming stable and unstable manifolds,see below).
For the geodesics directed as in Fig.19 it is intuitively clear that the flow
is expanding(unstable) .If we would change the direction of geodesics(i.e.if
we change the time evolution from t to -t),then the flow will be
contracting.It can be shown$^{18}$ that the rate of expansion is e while the
rate of contraction is e$^{-1}$ (e=exp).Now ,we are ready for a general
definition of the Anosov flow.

\textbf{Definition} \textbf{4.2}.\textit{If A : M}$\rightarrow$\textit{M is
some diffeomorphism of compact manifold M such that the tangent bundle T}%
$_{x}$\textit{M for any point x }$\in$\textit{M is decomposable into direct
sum }$\oplus$\textit{\ of the type}

\quad%
\begin{equation}
\text{T}_{x}\text{M=X}_{x}\oplus\text{Y}_{x} \tag{4.7}%
\end{equation}
\quad\quad\quad\textit{so that in some (Riemannian) metric ,and for some
number }$\lambda>1$\textit{\ the following inequalities hold}
\begin{equation}
\left\|  A_{*}\xi\right\|  \geq\lambda\left\|  \xi\right\|  \text{ \quad
}\forall\xi\in\text{X}_{x}\text{ \quad and \quad}\left\|  A_{*}\eta\right\|
\leq\lambda^{-1}\left\|  \eta\right\|  \text{\quad}\forall\eta\in\text{Y}_{x}
\tag{4.8}%
\end{equation}
\textit{where }$A_{*}$\textit{\ is the differential of the operator }%
$A$\textit{\ (acting in tangent space)and \quad}$\left\|  ...\right\|
$\textit{\ is the usual square of the length in some (Riemannian) metric
space,then the flow }T$_{x}$M\textit{\ is Anosov. \quad}\quad\quad\quad
\quad\quad\quad\quad\quad\quad\quad\quad\quad\quad\quad\quad\quad\quad
\quad\quad\quad\quad\quad\quad\quad\quad\quad\quad\quad\quad\quad\quad
\quad\quad\quad\quad\quad\quad\quad\quad\quad

\textbf{Remark} \textbf{4.1}. \textit{If ,in addition,the flow contains some
singularities,then it is called pseudo-Anosov}$[44]$\textit{.The sets X}$_{x}
$\textit{\ and Y}$_{x}$\textit{\ are called (un)stable foliations
respectively}.

\textbf{Remark} \textbf{4.2.}\textit{\ It can be shown,that the geodesic flow
on H is ergodic}$[18]$\textit{\ .That is there is some measure }$\mu(x)
$\textit{\ such that}%

\begin{equation}
\int d\mu(x)\varphi(x)=\lim_{T\rightarrow\infty}\frac{1}{T}\int\limits_{0}%
^{T}dt\varphi(A^{t}x(t)) \tag{4.9}%
\end{equation}
\textit{where }$A^{t}$\textit{\ is some evolution operator (continuous analog
of }$A_{*}$\textit{\ in (4.8)}).

In the case of pseudo-Anosov flows it is customary to introduce the transverse
(vertical) measure which can be intuitively understood using the following
picture,e.g.see Fig.21\marginpar{Fig.21} .That is, if \textbf{locally} the
foliation looks like the set of horizontal lines(away from
singularities),Fig.21a),then, when we try to restore the global picture,the
local charts should be glued in such a way that the vertical distance $\Delta$
between the leaves on one chart is in agreement with the vertical distance
$\Delta$ on another. If this \textbf{is }the case,the foliation is called
\textbf{measured }.We shall be dealing only with measured foliations $[10,25]$
in the rest of this work in accord with Poenaru$[10]$ and Langevin $[11]$ .

The above nice picture leaves us with no clues about the mechanisms by which
the flows can be changed from the reducible to periodic or from the reducible
to pseudo-Anosov,etc. As in the case of circle homeomorphisms (section
3),there must be some \textbf{physical }reasons of changing from one regime to
another.The best way to get some physical feeling of the processes which cause
''phase transitions''(i.e.changes in the flow regime) is through the notion of
''\textbf{train} \textbf{tracks}'' introduced by Thurston $[20]$ and
subsequently developed by many other mathematicians,e.g.see Ref.[45].The idea
behind the train track is simple but very powerful. To appreciate it , it is
helpful to recognize that there is some similarity between the way knots(or
links) are described and the train tracks. In the case of knots (links) one
usually studies a ''shadow'' of a knot obtained by projecting it onto some
arbitrary plane$[46]$.The projection is a four-valent planar graph K without
the dangling (''dead'') ends.Evidently,different knots may have \textbf{the
same} projection for a plane which orientation is fixed. To distinguish
between different knots(links) one has to resolve each 4-valent vertex (i.e.
to decide which strand is ''over'' and which is ''under'') and,in addition,to
use the set of isotopy moves (the Reidemeister moves) to bring one projection
in accord with the other.It is believed,that ,at least in principle,one can
decide if two knots are equivalent by performing some finite (but could be
very large !)sequence of the Reidemeister moves. In all these operations the
physical nature of a knot is not playing any role.In the case of train tracks
one is also dealing with a graph T without dead ends and also there are the
isotopy moves(analogous to the Reidemeister moves), but,in addition,
\textbf{and this is the most important},there are moves which \textbf{do not}
respect isotopy.These moves change the topology of the graph T and could be
associated with some physical processes as we shall explain below and in Part
II (section 6 and Appendix).The graph T has two basic building blocks which
can be easily recognized,e.g. see Fig.22\marginpar{Fig.22} .One can see from
this figure that,instead of having all leaves and precise angles ,one can
actually ''survive'' only with topologically equivalent objects which are
smooth at joints.With the basic building blocks just defined ,we can construct
our first train track.In particular,instead of having rather complicated
foliation pattern depicted in Figs.13,14,we can draw the following graph
depicted in Fig.23. \marginpar{Fig.23}Of course,the r.h.s.of Fig.23 provides
still only a coarse grained foliation pattern thus creating an illusion that
the line is closed (and this is indeed the case for the ''rational''
labyrinth).The numbers on the graph are quite arbitrary but are subject to one
restriction:at every switch the sum of ''entering'' numbers is equal to the
sum of the ''exiting '' numbers .The numbers are associated with the invariant
transverse measures which were formally introduced earlier. Since one can
construct from the collection of 3-valent vertices the vertex of more
complicated nature ,e.g. 4-vertex ,5-vertex,etc., it makes sense to talk about
given vertex v$_{k}$ (k=1-M) in general.. If \{$a_{i}^{in}\}$ is the set of
input branches and \{$a_{j}^{out}\}$ is the set of output branches ,then for a
given vertex v$_{k}$ the \textbf{switch condition} can be written as:%

\begin{equation}
\sum\limits_{i}\mu_{\text{v}_{k}}(a_{i}^{in})=\sum_{i}\mu_{\text{v}_{k}}%
(a_{i}^{out})\text{ .} \tag{4.10}%
\end{equation}
The switch condition is analogous to the Kirkhoff rule for currents known in
physics [$47]$ .It could be proven $[20,44]$ ,that the assignment of weights
(transverse measures) $\mu(a_{i})$ for each branch of the train track which is
subject to the switch conditions ,Eq.(4.10), at each vertex is equivalent to
the reconstruction of the entire measured foliation (up to isotopies and
Whitehead moves,Fig.4) [23]. The new feature which makes the train tracks more
complicated than knots (links) lies in \textbf{additional moves }which,unlike
the isotopy,do \textbf{change the topological type} of the track.These moves
have the major physical significance as it will be demonstrated shortly.Before
doing so ,we would like to provide a list of topology changing moves .They are
depicted in Fig.24.\marginpar{Fig.24} The shift operation can be performed
\textbf{without }reference to weights but,of course,the switch condition,
Eq.(4.10),should be obeyed before and after the shift.The split (or collapse)
is dependent ,however,upon the particular distribution of weights.The shift
does not destroy (or create) the existing singularities while the split
(collapse) may destroy (create) the singularities thus apparently violating
the P-H theorem.This deficiency can be easily corrected if some physics is
taken into consideration .To do so,several steps are still
required.\textbf{First}, following Ref.[45] ,let us consider the process of
collapse as the matrix operation. That is, we form the column vector made of
weights $a^{\prime},b^{\prime},...,e^{\prime}$ and look for the transition
matrix $\mathbf{S}^{-1}$ such that $\mathbf{x=S}^{-1}\mathbf{x}^{\prime}$
or,more explicitly,when $c^{\prime}<a^{\prime}$ ,we obtain,%

\begin{equation}
\left(
\begin{array}
[c]{l}%
a\\
b\\
c\\
d\\
e
\end{array}
\right)  =\left(
\begin{array}
[c]{lllll}%
1 & 0 & 0 & 0 & 0\\
0 & 1 & 0 & 0 & 0\\
0 & 0 & 1 & 0 & 0\\
0 & 0 & 0 & 1 & 0\\
1 & 0 & 0 & 1 & 1
\end{array}
\right)  \left(
\begin{array}
[c]{c}%
a^{\prime}\\
b^{\prime}\\
c^{\prime}\\
d^{\prime}\\
e^{\prime}%
\end{array}
\right)  \tag{4.11}%
\end{equation}
This result can be easily understood based on Eq. (4.10) and Fig. 24.
Indeed,according to Eq. (4.10) we have $a^{\prime}+e^{\prime}=c^{\prime}$
($c^{\prime}>a^{\prime})$ and also $e^{\prime}+d^{\prime}=b^{\prime}$ for the
split diagram while $a+b=e$ and $c+d=e$ for the collapsed diagram. At the same
time, using Eq.(4.11), we obtain: $a=a^{\prime};b=b^{\prime};c=c^{\prime
};d=d^{\prime}$ but $e=a^{\prime}+d^{\prime}+e^{\prime}.$The last result
becomes an identity if the Kirkhoff sum rules ,just obtained, are utilized.
Evidently,the corresponding matrices can be constructed for all elementary
processes depicted in Fig.24 and,furthermore,if ,for example, $\mathbf{S}$ is
the split matrix, then $\mathbf{S}^{-1}$ is the corresponding collapse
matrix,etc. \textbf{Second}, instead of considering the individual (local)
processes ,we can consider the state vector of the \textbf{entire }graph T
which is just a column vector of \textbf{all} transverse measures which we
denote as $\vec{\mu}$ .The evolution of the entire graph is determined then by
some transition matrix $\mathbf{T}$ so that%

\begin{equation}
\vec{\mu}^{\prime}=\mathbf{T}\vec{\mu} \tag{4.12}%
\end{equation}
\textbf{Third },this equation is still unphysical.To make it physical,we have
to ascribe certain statistical weights to \textbf{each} of the diagrams
depicted in Fig.24. The weights could be chosen either on the basis of
statistical mechanics arguments (e.g. in the case of liquid crystals)or on the
basis of quantum mechanical arguments ( e.g. in the case of gravity). For
instance,in the case of liquid crystals one can think of energy of the defect
formation and destruction (for more details ,please,see the next section and
sections 4,6 of Part II)as well as about the average energy (per defect) as a
function of $\nu$ (defined by Eq.(2.5)).Irrespective of the specific form of
the weights ,it is clear ,that the time evolution of the graph T can be
described in terms of the master equation(as it is usually done in statistical mechanics)%

\begin{equation}
\frac{\partial\mu_{i}}{\partial t}=\sum\limits_{i\neq j}(W_{ij}\mu_{j}%
-W_{ji}\mu_{i})\text{ .} \tag{4.13}%
\end{equation}
The matrix $W_{ij}$ is \textbf{not }symmetric unfortunately.This can be seen
already from the Eq.(4.11).In the case of 2+1 gravity the above equation could
,in principle,provide an alternative to the Wheeler-DeWitt equation$[48]$
.Eq.(4.13) may have time -independent solution which (in terms of discrete
maps,e.g. see section 3) ,by analogy with Eq.(4.5) ,we may call
\textbf{reducible} .It may as well have a periodic (in time )solution which we
may (or may not) associate with the previously defined \textbf{periodic(}in
space)\textbf{\ }case. Finally, it may not have any stationary or periodic
solutions.This then will be indicative of the pseudo-Anosov type of
evolution.The actual analysis of Eq.(4.13) would require a large scale
numerical work and,evidently,some drastic approximations will be
required.Therefore,for the time being,we would like to explore yet another possibility.

Before doing so,several remarks should be made.

\textbf{Remark} \textbf{4.3}. \textit{Since in the case of train tracks there
are no dead ends,this concept should be applied to open surfaces,e.g.to the
open disc D}$^{2}$\textit{\ ,with some caution,e.g.see Ref.[49].}

\textbf{Remark} \textbf{4.4}.\textit{\ Since the surface of non-negative Euler
characteristic contains no train tracks[20,45] some care should be taken to
by-pass this difficulty.This is discussed to some extent in the Appendix (see
also Refs.[33] and [49])}

\textbf{Remark} \textbf{4.5}.\textit{Without use of the probabilities leading
to the master Eq.(4.13), different outcomes of iteration had been studied
already }$[45,49,50]$\textit{\ .Clearly,the process is reducible ,periodic or
pseudo-Anosov depending upon what sequence of moves,Fig.24 ,had been used.For
example, if only direct processes are being considered,such as the right (R)
and the left (L) splits and the shift (S), then the matrix sequence could be
associated with some words in the alphabet made up of 3 letters }%
$[50]$\textit{so that ,in general,Eq.(4.13) describes a random walk on the
mapping class group }$M(R).$

Incidentally ,the\textbf{\ }3 letters\textbf{\ }alphabet is being used for the
meanders $[34]$. Hence,we would like now to study how one can use the meanders
in resolving the types of dynamical behaviour.This is accomplished to some
extent in the next section while in the Appendix we provide some evidence that
the meandritic labyrinths are indeed associated with the pseudo-Anosov homeomorphisms.

\section{Phase transitions in the system of meandritic labyrinths}

Phase transitions in the system of meanders was recently considered in
Ref.[35] in connection with the problem of folding of polymers.At the same
time,the arguments used in Ref.[35] could be traced back to the arguments made
by Peierls in connection with phase transintion in 2 dimensional Ising model
[51] .To facilitate reader's understanding of the development which follows
,we would like to provide here a summary of the arguments made by Peierls.

In the case of planar Ising model one is expecting to have droplets of spins
of the ''wrong'' sign embedded among the ''bulk'' which has the ''correct''
sign (e.g. ''+'' for the correct and ''-'' for the wrong).The interface
between the correct and the wrong spins determines the interfacial energy
E$_{L}$ of the droplet so that for the droplet of perimeter L the energy is
given by E$_{L}=LJ$ with $J$ being known ,in principle,constant.The partition
function Z$_{I}$ for the Ising model can now be written as follows,%

\begin{equation}
Z_{I}=2\sum\limits_{L=0}^{\infty}G_{L}\exp\{-\beta JL\}\text{ ,\quad}%
\beta^{-1}=k_{B}T, \tag{5.1}%
\end{equation}
where $k_{B}T$ is the usual temperature factor.Since the Ising model is
defined on the lattice ,$L$ is necessarily discrete and $G_{L}$ is the
combinatorial entropic factor which determines the number of ways the droplets
of the total perimeter length $L$ can be arranged on the lattice.

The difficulty in calculating Z$_{I}$ lies in determining the factor $G_{L}$
correctly .If, following Peierls,we make a drastic approximation :$G_{L}%
\simeq4^{L}($ since for the square lattice the coordination number is 4),then
Z$_{I}$ can be calculated at once with the result%

\begin{equation}
Z_{I}\simeq(1-4\lambda)^{-1} \tag{5.2}%
\end{equation}
,where $\lambda=\exp\{-\beta J\}.$ The above expression makes sense only for
$\lambda<0.25$ so that 0.25 determines the critical temperature of the Ising
model through equation : 4$\lambda^{\ast}=1$ .This result is interesting to
compare against the exact result $[52]$ :$\lambda^{\ast}=\sqrt{\sqrt{2}-1}$
$\simeq0.64$ (for $J=1)$ .Obviously,the value 0.25 appears to be too low even
if compared with the Bethe approximation$[53]$ which yields 3 $\left(
\lambda^{\ast}\right)  ^{2}=1$ .This was noticed already by Peierls $^{51}$
.Nevertheless,the above crude estimate can be systematically improved and the
Peierls arguments are the most powerful tool in general study of phase
transitions in discrete spin systems as is well known $[53]$ .The exact result
,$\left(  \lambda^{\ast}\right)  ^{2}=\sqrt{2}-1,$ is valid only for the
square lattice.In the case of triangular lattice the exact result is
different:$\left(  \lambda^{\ast}\right)  ^{2}=\left(  \sqrt{3}\right)  ^{-1}%
$,while for the hexagonal lattice ,it is different again : $\left(
\lambda^{\ast}\right)  ^{2}=(2+\sqrt{3})^{-1},$ e.g.see Ref.[52].Hence,the
critical temperature is actually a function of two variables :$J$ and z ,where
z is the coordination number of the lattice.Something similar will happen in
our (meandritic )case too.

In the case of Ising model we can effectively define the notion of an order
parameter.This is less obvious in the case which we are going to consider.
What is important for us,however,is the fact that the partition function
Z$_{I}$ diverges.We shall ,by analogy with the Ising model ,associate such
divergence with the phase transition,in our case from the
meandritic(pseudo-Anosov,according to Ref.[33] and Appendix) to the periodic
and/or reduced phases (since these are the only existing possibilities).

Using the results of section 2 ,Eq.(2.5) and section 4 of Part
II,Eq.(4.49),the total surface energy of n defects can be estimated as%

\begin{equation}
E_{n}\geq\frac{\tilde{K}}{4}nq^{2}\ln\left(  \frac{1}{\nu}\right)  \equiv
n\hat{J} \tag{5.3}%
\end{equation}
where the charge q is equal to the index of the singularity,Eq.(1.11),and
$\tilde{K}$ is some known constant (defined in section 4 , Part II) related to
the surface tension. In arriving at Eq.(5.3) we had assumed that only Y's and
thorns are present (so that the magnitude of charges are all the same).To
estimate the combinatorial factor (analogous to $G_{L}$ in Eq.(5.1)) the
following arguments are helpful.

According to Ref.[36] ,any meander can be built by superposition of two arc
configurations of the same order :one is considered to be the top while
another is the bottom as depicted in Fig.25.\marginpar{Fig.25.}Since both the
top and the bottom are configurations of the same order,then by concentrating
our attention,say ,on the top configuration,we can obtain a projective meander
(e.g. see section 3 )by identifying the beginning and the end of the
horizontal straight line.In this case we already know that the number of
possible configurations is $C_{n}$ .Evidently,we can do the same for the
bottom arc system thus obtaining another $C_{n}.$ If we try all possible
permutations,then,a priory,there is no guarantee that if the top and the
bottom are connected together we shall obtain just one connected meander.More
likely ,we shall obtain $C_{n}^{2}$ meanders some of them connected and some
not.More exactly, let M(x) be the meander generating function,i.e.%

\begin{equation}
\text{M(x)=}\sum\limits_{n=0}^{\infty}\text{M}_{n}\text{x}^{n}, \tag{5.4}%
\end{equation}
where M$_{n}$ was defined in section 3.There is the following

\textbf{Theorem} \textbf{5.1}. \quad\quad\quad\quad$C_{n}\leq$M$_{n}\leq
C_{n}^{2}$

\textbf{Proof} .The upper bound was just established.The lower bound can be
established if we associate the system of meanders with the 3 letter alphabet
(as was briefly mentioned in section 4).For more details,please,consult
Ref.[34]. $\square$

Among all of $C_{n}^{2}$ meanders of order n \textbf{not }all are
topologically different.Let N$_{n}$ denote the number of topologically
distinct configurations of meanders which altogether pass through 2n points
(which actually may belong to different lines),then the generating function
for these numbers could be defined as%

\begin{equation}
\text{N(x)=}\sum\limits_{n=0}^{\infty}\text{N}_{n}\text{x}^{n}. \tag{5.5}%
\end{equation}
If we disrespect the topological differences,then the generating function for
\textbf{the system of meanders} can be written as%

\begin{equation}
\text{B(x)=}\sum\limits_{n=0}^{\infty}C_{n}^{2}\text{x}^{n}. \tag{5.6}%
\end{equation}

\textbf{Theorem 5.2}. \textit{Functions B(x) and N(x) are connected with each
other through the functional equation}%

\begin{equation}
\text{B(x)=N(xB}^{2}\text{(x))} \tag{5.7}%
\end{equation}

\textbf{Proof}.Please,consult Ref.[34]. $\square$

\textbf{Corollary} \textbf{\-5.2} .\textit{The meandritic numbers M}%
$_{n}^{\left(  k\right)  }$\textit{\ introduced in section 3 are generated via
the following generating function }$[35,36]$%

\begin{equation}
\text{m}_{n}\text{(g)=}\sum\limits_{k=0}^{n}\text{M}_{n}^{\left(  k\right)
}\text{g}^{k} \tag{5.8}%
\end{equation}
\textit{provided that m}$_{n}$\textit{(g=1)=}$C_{n}^{2}.$

Based on the results just described,we are ready now to write down the
partition function for the system of meandritic labyrinths. It is given by%

\begin{equation}
Z_{g}(\text{x})=\sum\limits_{n=0}^{\infty}\sum\limits_{k=0}^{n}\text{M}%
_{n}^{\left(  k\right)  }\text{g}^{k}\text{x}^{n}\text{ \quad, x=exp\{-}%
\beta\hat{J}\}, \tag{5.9}%
\end{equation}
where $\hat{J}$ is defined by Eq.(5.3).This equation ,in principle,is exact
but,it requires knowlege of g.Evidently,g is the fugacity which determines the
average number of meanders in the cluster $<$k$>$ which is given by%

\begin{equation}
<\text{k}>=\text{g}\frac{\partial}{\partial\text{g}}\ln Z_{\text{g}}(\text{x)
.} \tag{5.10}%
\end{equation}
This quantity is hard to estimate (see,however ,some attempts in this
direction below).And,even if we would succeed,still,we would need to invert
the infinite power series in order to write the fugacity g in terms of
$<$k$>.$ Therefore,we would like to make an approximation based on the exact
result coming from Eq. (5.8) for g=1.Thus ,we obtain the following
approximation for the partition function of the system of meanders:%

\begin{equation}
\text{Z(x)=}\sum\limits_{n=0}^{\infty}C_{n}^{2}\text{x}^{n}\text{ \quad, \quad
x=exp\{-}\beta\hat{J}\}. \tag{5.11}%
\end{equation}
Remarkably enough,this partition function admits the \textbf{exact
}resummation$[34]$ with the result (t$^{2}=$x) :%

\begin{equation}
\text{Z(t}^{2})=\frac{1}{4\text{t}^{2}}\left(  -1+\frac{1}{2\pi}%
\int\limits_{0}^{2\pi}d\varphi\sqrt{\text{ }1-8\text{t}\cos\varphi
+16\text{t}^{2}}\right)  . \tag{5.12}%
\end{equation}
At the same time,one can study the convergence of Z(x) based on known
asymptotic value for $C_{n}:$%
\begin{equation}
C_{n}\simeq const\frac{4^{n}}{n^{\frac{3}{2}}}\text{ \quad, }n\rightarrow
\infty. \tag{5.13}%
\end{equation}
From here,we obtain,%

\begin{equation}
\frac{C_{n+1}^{2}}{C_{n}^{2}}\sim16 \tag{5.14}%
\end{equation}
and,therefore,0$<$x$\leq\frac{1}{16}$ .Substitution of x$^{*}=16$ into
Eq.(5.12) produces \textbf{finite }result :%

\begin{equation}
\frac{1}{4}\text{Z(}\frac{1}{16})=\left(  \frac{4-\pi}{\pi}\right)  \text{ .}
\tag{5.15}%
\end{equation}
For x$>\frac{1}{16}$ the partition function diverges,of course.

Clearly,the condition 16x$^{*}=1$ determines the critical temperature for the
fixed value of $\nu$ or determines the critical ''density'' $\nu^{*}$ for a
fixed temperature and the value of surface tension \~{K} in view of Eq.(5.3).

The analysis presented above is still incomplete since one can calculate now
N(x) in view of Eq.(5.7). Following Ref.[34] ,let us multiply both sides of
eq.(5.7) by t (t$^{2}$=x).We obtain,%

\begin{equation}
\Phi(\text{t)=tB(t}^{2})=\text{tN(}\Phi^{2}(\text{t)) .} \tag{5.16}%
\end{equation}
Let,furthermore, $\lambda=\Phi($t) so that t=$\Phi^{-1}(\lambda).$ Then,we obtain,%

\begin{equation}
\text{N(}\lambda^{2})=\frac{\lambda}{\Phi^{-1}(\lambda)}\text{ .} \tag{5.17}%
\end{equation}
From here it follows,that since using closed form result , Eq.(5.12),$\Phi
($t)=tZ(t$^{2}$) can be calculated,at least in principle, then N($\lambda^{2})
$ \quad can be calculated as well.The radius of convergence for N(x) is equal
to $\left(  \frac{4-\pi}{\pi}\right)  ^{2}$ .Obtained results are in formal
qualitative accord with that known in physical literature on liquid crystals
$[14].$ But now we know that the disordered (liquid-like) phase is actually
pseudo-Anosov while the ordered (hexatic) is periodic and the reducible is
solid-like. The phase transition mechanism ,however,has absolutely nothing to
do with the Kosterlitz-Thouless type of transition which was discussed in
section 2. Additional details related to this subject are provided in section
4 ,Part II.

\textbf{Note added in proof}: When this paper was completed we had found a
paper by Ishikawa and Lawrentovich (Europhys. Lett. \textbf{41},pp.171-176
(1998)) in which the Whitehead moves depicted in our Fig.4 had been detected
experimentally. In addition, the paper by Penner (Adv.in Math.\textbf{101}%
,pp31-49 (1993)) had also came to our attention. In this paper the
train-tracks are being used to describe folding of RNA. Since in Ref.[35] the
folding of proteins was considered with help of the meanders, the above
references provide additional support to the results presented in sections 3-5
which relate meanders to train tracks. More details could also be found in
section 7.5 of Part II.

\mathstrut

\quad\quad\quad\quad\quad\quad\quad\quad\quad\quad\quad\quad\textbf{Appendix}

\mathstrut

In this Appendix we would like to provide some arguments in favor of the
statement made in section 4 that the meandritic labyrinths are associated with
the pseudo-Anosov homeomorphisms.

To this purpose let us consider the simplest labyrinth which is made out of
two copies of the disc D$^{2}$ both of which containing just two thorns (to be
in accord with P-H theorem).When glued properly,these two discs will form a
sphere S$^{2}$ with foliation forming a labyrinth (e.g.see Ref.[33],page
29).We need now to explain why such foliation could be of pseudo-Anosov type.

Consider an auxiliary problem about the foliations on the torus$[17,18,41]$
.If we regard the torus T$^{2}$ as a quotient of \textbf{R}$^{2}$ by the
integer lattice Z$^{2}$ ,then the homeomorphisms $h_{\alpha}$ of T$^{2}$ are
generated by the group GL$_{2}(Z)$ since any element $\alpha$ of GL$_{2}(Z)$
maps Z$^{2}$ into itself thus inducing the continuous map $h_{\alpha}:$
T$^{2}\rightarrow$T$^{2}$ .The homeomorphism is orientation preserving if
\quad det ($\alpha)=1$ ($\forall\alpha\in$GL$_{2}(Z))$ and,in this case
,GL$_{2}(Z)$ becomes SL$_{2}(Z)$ which is represented by 2$\times2$ matrix
given by
\begin{equation}
\left(
\begin{array}
[c]{cc}%
a & b\\
c & d
\end{array}
\right)  \in\text{SL}_{2}(Z)\text{ \quad if }ad-cb\text{=1.} \tag{A.1}%
\end{equation}
The eigenvalues of the above matrix are obtained, as usual, as the roots of
the characteristic polynomial%

\[
t^{2}-(a+d)t+(ad-cb)=0
\]
or,in view of (A.1) ,%

\begin{equation}
t^{2}-(a+d)t+1=0. \tag{A.2}%
\end{equation}
There are three possibilities in general:

1)the roots of (A.2) are complex (when a+d=0,1 or -1);

2)both roots of (A.2) are equal to $\pm1$ (when a+d=2 );

3)the roots of (A.2) are distinct reals (when $\left|  a+d\right|  >2$) .\quad\ \ \ \ \ \ \ \ \ \ \ \ \ \ \ \ \ \ \ \ \ \ \ \ \ \ \ \ \ \ \ \ \ \ \ \ \ \ \ 

\ \ By analogy with the circle maps considered in section 3,it can be
shown$[16,41]$,that the first possibility produces \textbf{periodic }maps of
T$^{2} $ ,the second produces \textbf{reducible }maps of T$^{2}($that leave a
simple Jordan curve fixed (possibly with reversed orientation))while the third
possibility is responsible for the Anosov (\textbf{not pseudo-Anosov)} type of
flow $[17,18]$ on T$^{2}$ which was defined in section 4.Consider now our
S$^{2}$ ,made up of two copies of D$^{2}$,each containing two thorns.
According to the results of Part II (section 3) thorns are singularities in
the complex z-plane (or on S$^{2}$ ) of the type $\left(  \sqrt{z}\right)
^{-1}$ .As it is usually done in the complex analysis,we can make branch cuts
(as depicted in Fig.A1)\marginpar{Fig.A.1}in order to construct the two
sheeted Riemann surface.In Fig.A.1 these two sheets are depicted as onion-like
sphere with two layers.If the outer sphere is peeled off and placed opposite
to the inner sphere,as depicted in Fig.A.2, then,by obvious identification of
edges ,the surface which is homeomorphic to T$^{2}$ is formed.Hence,T$^{2}$ is
a two-fold ramified branched covering of S$^{2}$ .The points of ramification
are the above thorn-like singularities.In general,one can prove the following

\textbf{Theorem A.1}. \textit{Every closed orientable surface (of finite
genus) is two-fold ramified branched cover of S}$^{2}$\textit{\ .}

\textbf{Proof} .Please,consult Refs.[55-57]. $\square$ . \quad\quad\quad
\quad\quad\quad\quad\quad\quad\quad\quad\quad\quad\quad\quad\quad\quad
\quad\quad\quad\quad\quad\quad\quad

Finally, if for T$^{2}$ we have the case of the Anosov flow,then it is being
transferred onto S$^{2}$ in the form of pseudo-Anosov flow (because of 4
singularities placed on S$^{2}$ ).More exactly,the foliation on S$^{2}$ is the
quotient of the Anosov homeo(or diffeo)morphism on T$^{2}$ via 2-fold ramified
branched covering of S$^{2}$ by T$^{2}$ . For more than 4 defects in view of
Theorems A.1 and 4.3 of the main text ,it is also possible ,in principle,to
establish connection between the labyrinths and pseudo-Anosov
homeomorphisms.Please,consult Expos$e\prime$11 and Expos$e\prime$12 of Ref[23]
and section 4 of Ref.[44] for additional details.For more information on
branched coverings,please,consult Ref.[57].

\newpage

\mathstrut

\quad\quad\quad\quad\quad\quad\quad\quad\quad\quad\quad\quad\textbf{Figure Captions}

\medskip

Fig.1 Typical singularities of the vector fields in the plane

Fig.2 Typical singularities of the line field

Fig.3 The field lines around the dipole a) and the source b)

(In the case of a sink the directions of lines are changed into the opposite)

Fig.4 The allowed Whitehead moves for the line fields(''creation'')

Fig.5 The additional configurations leading to ''destruction''of defects

Fig.6 Possible ''periodic'' phase in the case of minimal number of line field

defects in the disc

Fig.7 Possible ''periodic ''phase made up of 14 defects in the disc.

Fig.8 Possible hexatic-type phase obtained in the thermodynamic limit

Fig.9 The elementary triangle used in P-H theorem calculations related

to the hexatic-type phase depicted in Fig.8

Fig.10 a)Another possible hexatic phase which is permitted by the Euler

theorem;

b)The elementary triangle used in Hopf-like calculations of the

Euler characteristic

Fig.11 Two dimensional plane can be covered by hexagons.At the same

time,one can cover the plane by the triangular lattice so that for the

complete coverage 3 hexagons should fit together around each vertex

Fig.12 Simplest ''building blocks'' for measured foliations

Fig.13 Simplest nontrivial global foliation pattern made of simple

building blocks

Fig.14 The structure of disordered phase (labyrinth) in the case

of minimal number of line defects (to be compared with the

ordered phase depicted in Fig.6)

Fig.15 The simplest example of the meander labyrinth

Fig.16 The simplest meanders of order n=1-3

Fig.17 Geodesics on the Poincare upper H- plane

Fig.18 Geodesics on the unit open disc D$^{2}$

Fig.19 Anosov flow on H-plane

Fig.20 Anosov flow on the disc D$^{2}$

Fig.21 For the measured foliation the vertical distance between the

leaves on two adjacent patches (local charts) must be the same

Fig.22 Basic building blocks of the train track

Fig.23 A simple train track and the foliation pattern which is

associated with it

Fig.24 Major topology nonpreserving moves for the train tracks

Fig.25 Construction of a typical meander

Fig.A.1 Two sheeted covering of S$^{2}$ before glueing

Fig.A.2 Identification of cuts between the inner and the outer spheres

produces torus T$^{2}$

\newpage

\mathstrut

\quad\quad\quad\quad\quad\quad\quad\quad\quad\quad\quad\quad\quad\textbf{References}

[1] M.Nakanara,\textit{\ Geometry,Topology and Physics },

Adam Hilger, Bristol,1990 .

[2] J.Moore,\textit{\ Lectures on Seiberg-Witten Invariants ,}

Springer-Verlag, Berlin ,1996

[3 ] G.Hemion, \textit{The Classification of Knots and 3-Dimensional Spaces },

Oxford U.Press,Oxford ,1992.

[4 ] H.Hopf,\textit{\ Differential Geometry in the Large ,} LNM No.1000,

Springer-Verlag, Berlin,1989 .

[5] V.Nemytskii and V.Stepanov ,\textit{\ Qualitative Theory of Differential}

\textit{\ Equations, }Princeton U.Press , Princeton,1960.

[6] G.Godbillon, \textit{Dynamical Systems on Surfaces ,}Springer-Verlag,

Berlin,1983.

[7] A.Kholodenko,\textit{\ Use of Quadratic Differentials for Description }

\textit{of Defects and Textures in Liquid Crystals and 2+1 }

\textit{Gravity } (to be submitted).

[8] C.Oseen, The theory of liquid crystals, Faraday

Transactions 29\textbf{(}1993\textbf{) }883 -899.

[9] F.Frank,On the theory of liquid crystals ,Faraday

Discussions 25(1958) 19- 28.

[10] V.Poenaru, Some aspects of the theory of defects in ordered

media and gauge fields related to foliations ,Comm.Math.Physics

80 (1981)127 -136 .

[11] \ R.Langevin, Foliations,energies and liquid crystals,

Asterisque107-108(1983)201-213 .

[12] S.Deser , R.Jackiw and G.'t Hooft ,Three dimensional Einstein

gravity:dynamics in flat space, Ann.Physics 152(1984)220-235 .

[13] P. Menotti and D.Seminara ,Energy theorem for \ 2+1

dimensional gravity,Ann.Phys.240(1995) 203-221.

[14] P.Chaikin and T.Lubensky,\textit{\ Principles of Condensed }$\mathit{M}%
$\textit{atter Physics, }

Cambridge U.Press,Cambridge ,1995 .

[15] M.Katanaev and I.Volovich ,Theory of defects in solids and

three dimensional gravity, Ann.Phys.216(1992)1-28 .

[16] B.Halperin and D.Nelson, Theory of two dimensional melting,

Phys.Rev.Lett. 41(1978) 121-124 .

[17]V.Arnold ,\textit{Geometrical Methods in the Theory of Ordinary }

\textit{Differential Equations ,} Springer-Verlag, Berlin,1983 .

[18]A.Katok and B.Hasselblat, \textit{Introduction to the Modern Theory }

\textit{of Dynamical Systems} , Cambridge U.Press.Cambridge ,1997.

[19] V.Arnold, Branching covering CP$^{2}\rightarrow$S$^{4}$, hyperbolicity and

projective topology ,Siberian Math.Journal 29(1988)36-47 .

[20] W.Thurston,\textit{\ The Geometry and Topology of Three Manifolds} ,

Lecture Notes (Princeton U.,1979) http://www.msri.org/gt3m/

[21] E.Hague and P.Hemmer , The two dimensional Coulomb gas,Physica

Norvegica 5(1971)\textbf{\ },209-217. .

[22] J.Kosterlitz and D.Thouless,Metastability and phase transitions

in two dimensional systems ,J.Phys.C6(1973)1181-1203.

[23] A.Fathi,F.Laudenbach and V.Poenaru, Asterisque 66-67\textbf{\ }(1979).

[24] J-P Otal, Le theoreme d'hyperbolisation pour les varietes de

dimenson 3 , Asterisque 235\textbf{\ }(1996).

[25].W.Thusrton , On the geometry and dynamics of diffeomorphisms

of surfaces,Bull.Am.Math.Soc.19(1988) 417-432 .

[26] P.-G.de Gennes ,\textit{The Physics of Liquid Crystals },Clarendon Press,

Oxford,1979.

[27] \ L.Kinsey,\textit{\ Topology of Surfaces }, Springer-Verlag, Berlin,1993 .

[28] B.Halperin and D.Nelson,Errata, Phys.Rev.Lett.41\textbf{(}1978\textbf{)}%
519 and Ref.16.

[29] R.Baxter,\textit{\ Exactly Solved Models in Statistical Mechanics }

\textit{,}Academic Press, N.Y.1982 ;R.Monasson and O.Pouliquen, Entropy

of particle packings: an illustration on a toy model,

Physica A236(1997)395 -410.

[30] A.Kholodenko and Th.Vilgis ,The tube diameter in polymer melts,its

existence and its relation to the quantum Hall effect,

J.Phys.I(France) 4(1994) 843-862 .

[31] A.Kholodenko and Th.Vilgis , \textit{Some Geometrical and Topological }

\textit{Problems in Polymer Physics} , Physics Reports 298(1998) 251-372 .

[32] W.Thurston,\textit{Three-Dimensional Geometry and Topology },Vol.1,

Princeton.U.Press , Princeton,1997.

[33] H.Rosenberg,Labyrinths in the disc and surfaces,

Ann.Math. 117(1983)1-32.

[34] S.Lando and A.Zvonkin ,Meanders ,Selecta Mathematica

Sovetica 11(1992)117 -144.

[35] P.Di Francesco,O.Golinelli and E.Guitter,Meanders: a direct

enumeration approach, Nucl.Phys.B482 (1996) 497-535.

[36] P.Di Francesco,O.Golinelli and E.Guttier,Meanders and the

Temperley-Lieb algebra, Comm.Math.Phys.186(1997)1-59 .

[37] S.Lando and A.Zvonkin ,Plane and projective meanders ,

Theor.Computer Sci.117(1993) 227 -241.

[38] W.Thurston , in \textit{Low Dimensional Topology and Kleinian Groups} ,

pp.91-112 ,Cambridge U.Press,Cambridge ,1986.

[39] W.Abikoff,\textit{\ The Real Analytic Theory of Teichmuller Space },Springer-

Verlag,Berlin,1989.

[40] P.Buser,\textit{\ Geometry and Spectra of Compact Riemann Surfaces },Birkhauser,

Boston,1992 .

[41] A.Casson and S.Bleiler,\textit{\ Automorphisms of Surfaces After Nielsen}

\textit{and Thurston },Cambridge U.Press,Cambridge,1988 .

[42] A.Marden and K.Strebel,in \textit{Differential geometry and Complex }

\textit{Analysis ,}pp. 195-204, Springer-Verlag, Berlin,1985 .

[43] J.Birman,\textit{\ Braids,Links and Mapping Class Group, }Princeton U.Press,

Princeton,1975.

[44]A.Marden and K.Strebel ,Pseudo-Anosov Teichm\"{u}ller mapping,

J.D'Analyse Mathematique 46 (1986)194-220.

[45] R.Penner and J. Harer,\textit{Combinatorics of Train Tracks, }Princeton

U.Press, Princeton,1992 ; R.Penner in \textit{Low Dimensional Topology and}

\textit{\ Kleinian Groups ,}pp.77-90, Cambidge U.Press,Cambridge, 1986 .

[46] L.Kauffman,\textit{\ On Knots },Princeton U.Press, Princeton,1987.

[47] R.Bott and M.Duffin,On the algebra of networks, Trans.Am.Math.

Soc.74\textbf{\ (}1953\textbf{) }99-109 .

[48] G.Esposito , \textit{Quantum Gravity, Quantum Cosmology and Lorentzian }

\textit{Geometries },Springer-Verlag, Berlin,1992.

[49] J.Los , Pseudo-Anosov maps and invariant train tracks in the disc:

a finite algorithm, Proc.London Math.Soc.66 (1993)\textbf{\ }400-430 .

[50] A.Papadopoulos and R.Penner , A characterization of pseudo-Anosov

foliations , Pac.J.of Math.130(1987)\textbf{\ }359 -377 .

[51] R.Peierls ,On Ising's model of ferromagnetism, Proc.Cambr.Phil.Soc.

32(1936)\textbf{\ }477-481.

[52]C. Itsykson and J.Drouffe ,\textit{Statistical Field Theory ,}Vol.1 ,

Cambridge U. Press,Cambridge,1989.

[53] Ya.Sinai ,\textit{Theory of Phase Transitions },Pergamon Press,Oxford,1982.

[54] D.Rolfsen,\textit{\ Knots and Links },Publish or Perish,Houston,TX,1990.

[55] R.Fenn,\textit{\ Techniques of Algebraic Topology },Cambridge U.Press,

Cambridge,1983.

[56] J.Stillwell,\textit{Classical Topology and Combinatorial Group }

\textit{Theory },Springer-Verlag, Berlin,1993 .

[57] I.Berstein and E.Edmonds ,On the construction of branched

coverings of low dimensional manifolds, Transactions Am.Math.Soc.

247(1979)87-124.
\end{document}